\def\fr#1#2{\hbox{${#1\over #2}$}}
\def\subsub#1{\medskip{\bf #1}}     \def\ra{\rightarrow}
\def\lra{\leftrightarrow}           
                  \def\pd{\partial}
\def\ort{\perp}                     
              
        \def\tgr{GR$_{\parallel}$}
\def\mb#1{\hbox{\boldmath$#1$}}     
\def\bT{\bar T}                     \def\bcL{{\bar{\cal L}}}
\def\bcH{{\bar{\cal H}}}            \def\bphi{\bar\phi}

\def\m{\mu}             \def\n{\nu}              \def\k{\kappa}
          \def\g{\gamma}           \def\d{\delta}
                     
\def\a{\alpha}          \def\b{\beta}            
\def\vphi{\varphi}      \def\ve{\varepsilon}
\def\r{\rho}                       \def\p{\pi}
\def\l{\lambda}         \def\o{\omega}           
\def\bn{\bar n}         \def\bi{{\bar\imath}}    \def\bk{{\bar k}}
       \def\bm{{\bar m}}        \def\bn{{\bar n}}
   
\def\cL{{\cal L}}       \def\cH{{\cal H}}        \def\cP{{\cal P}}
               
\def\hp{{\hat\pi}}            \def\tcH{\tilde\cH}
\documentclass[12pt]{article}
\usepackage{latexsym,equations}
\def\nn{\nonumber}
\def\be{\begin{equation}}             \def\ee{\end{equation}}
\def\bea{\begin{eqnarray} }           \def\eea{\end{eqnarray} }
\def\lab#1{\label{eq:#1}}             \def\eq#1{(\ref{eq:#1})}
\def\bsubeq{\begin{subequations}}     \def\esubeq{\end{subequations}}
\def\bitem{\begin{itemize}}           \def\eitem{\end{itemize}}
\def\bull{\rule[.45ex]{1ex}{.3ex}\hskip1.1ex}
\newcommand{\sect}[1]{\addtocounter{section}{1}
    \section*{\large\thesection.~#1}\setcounter{equation}{0}}

\newcommand{\ack}[1]{\section*{\large #1}}
\newcommand{\app}[1]{\addtocounter{section}{1}
    \section*{\large Appendix \thesection:~#1}\setcounter{equation}{0}}

\begin{document}
\title{\vskip-1.4cm\large\bf
       Gauge symmetries of the teleparallel theory of gravity}
\author{\normalsize M. Blagojevi\'c and M. Vasili\'c\thanks{
                    Email addresses: mb@phy.bg.ac.yu;
                    mvasilic@phy.bg.ac.yu} \\
        \small {\it Institute of Physics, 11001 Belgrade,
                    P. O. Box 57, Yugoslavia}}
\date{}
\maketitle
\begin{abstract}
We study gauge properties of the general teleparallel theory of
gravity, defined in the framework of Poincar\'e gauge theory. It is
found that the general theory is characterized by two kinds of gauge
symmetries: a specific gauge symmetry that acts on Lagrange
multipliers, and the standard Poincar\'e gauge symmetry. The canonical
generators of these symmetries are explicitly constructed and
investigated.
\end{abstract}

\sect{Introduction}

Modern developments in particle physics suggest the possibility that
gravity might be described by a geometric structure different from
Riemannian space of general relativity (GR). Such geometric structures
appear naturally in gauge theories of gravity, which represent a
promising framework for describing gravitational interaction \cite{1}.
Here, we focus our attention on the teleparallel description of
gravity in the framework of Poincar\'e gauge theory (PGT)
\cite{2,3,4}. Basic dynamical variables in PGT are the tetrad field
$b^k{_\m}$ and Lorentz connection $A^{ij}{_\m}$, and the corresponding
field strengths are geometrically identified with the torsion and the
curvature:
\bea
&&T^i{}_{\m\n}=\pd_\m b^i{_\n}+A^i{}_{s\m}b^s{_\n}-(\m\lra\n)\, ,\nn\\
&&R^{ij}{}_{\m\n}=\pd_\m A^{ij}{_\n}
                  +A^i{}_{s\m}A^{sj}{_\n}-(\m\lra\n)\, .\nn
\eea
General geometric structure of PGT corresponds to Riemann-Cartan space
$U_4$, defined by metric (or tetrad) and metric compatible connection.

The teleparallel or Weitzenb\"ock geometry $T_4$ is defined as a
special limit of PGT by the requirement of vanishing curvature:
\be
R^{ij}{}_{\m\n}(A)=0 \, .                                   \lab{1.1}
\ee
The teleparallel theory has been one of the most attractive
alternatives to GR \cite{5,6,7} until the work of Kopczy\'nsky
\cite{8}. He demonstrated the existence of a hidden gauge symmetry, and
concluded that the theory is inconsistent since the torsion tensor is
not completely determined by the field equations. Nester improved the
arguments of Kopczy\'nsky by showing that the unpredictable behaviour
of torsion occurs only for some very special solutions \cite{9}.

The canonical analysis of the teleparallel formulation of GR, performed
in Ref. \cite{10}, was aimed at clarifying these unusual properties of
the teleparallel theory. In the present paper, we continue this
investigation by a detailed study of all gauge symmetries of the
general teleparallel theory. The precise form of the gauge generators
obtained in this paper can be used to introduce the important concepts
of energy, momentum and other conserved charges, and understand the
related role of asymptotic conditions in the teleparallel theory of
gravity \cite{11,12}.

We begin our considerations in Sect. 2 by introducing the basic
Lagrangian and Hamiltonian properties of the teleparallel theory. In
Sect. 3, we construct the gauge generator of a specific symmetry,
called $\l$ symmetry, which is present in any teleparallel theory.
Then, in Sect. 4, we study a simple model in order to understand the
relation between the $\l$ and Poincar\'e gauge symmetries. Finally, in
Sect. 5, we construct the Poincar\'e gauge generator for a general
teleparallel theory. Section 6 is devoted to concluding remarks.

Our conventions are the same as in Refs. \cite{10,12}: the Latin
indices refer to local Lorentz frame, the Greek indices refer to the
coordinate frame; the first letters of both alphabets
$(a,b,c,\dots;\a,\b,\g,\dots)$ run over $1,2,3$, and the middle alphabet
letters $(i,j,k,\dots;\m,\n,\l,\dots)$ run over $0,1,2,3$;
$\eta_{ij}=(+,-,-,-)$, $\ve^{ijkl}$ is completely antisymmetric symbol
normalized by $\ve^{0123}=+1$, $\d=\d(x-x')$; the Hamiltonian $H$ and
its density $\cH$ are related by $H=\int d^3x\cH$.

\sect{Basic dynamical features of the teleparallel theory}

\subsub{Lagrangian dynamics.} Gravitational dynamics of the general
teleparallel theory, in the framework of PGT, is described by a class
of Lagrangians quadratic in the torsion,
\bea
&&\cL = b\cL_T + \l_{ij}{}^{\m\n}R^{ij}{}_{\m\n}\, ,\nn\\
&&\cL_T=a\bigl(AT_{ijk}T^{ijk}+BT_{ijk}T^{jik}+CT_{k}T^{k}\bigr)
       \equiv \b_{ijk}(T)T^{ijk} \, ,                       \lab{2.1}
\eea
where the Lagrange multipliers $\l_{ij}{}^{\m\n}$ ensure the
teleparallelism condition \eq{1.1}, $A,B$ and $C$ are free parameters,
$a=1/2\k$ ($\k$ = Einstein's gravitational constant), and
$T_k=T^m{}_{mk}$. Note that, following Ref. \cite{12},
$\l_{ij}{}^{\m\n}$ is assumed  to be a tensor density rather then a
tensor, which leads to a slightly simplified constraint analysis, as
compared to Ref. \cite{10}.

The physical relevance of the theory \eq{2.1} lies in the fact that
there is a one-parameter family of teleparallel theories, defined by
the conditions
\bitem
\item[$i)$] $2A+B+C=0$,\ $C=-1$,
\eitem which passes all the standard gravitational tests \cite{5,6,7};
hence, it is empirically indistinguishable from GR. One particularly
interesting member of this family is defined by
\bitem
\item[$ii)$] $2A-B=0\quad$ (i.e. $A=1/4$,\ $B=1/2$,\ $C=-1)$.
\eitem
Using a well known geometric identity \cite{10}, one finds that
this choice leads effectively (up to a four-divergence) to the
Hilbert-Einstein Lagrangian defined in Riemann space $V_4$. We call
this theory the teleparallel form of GR, and denote it by \tgr.

By varying the teleparallel Lagrangian \eq{2.1} with respect to
$b^i{_\m}$, $A^{ij}{_\m}$ and $\l_{ij}{}^{\m\n}$, we obtain the
following gravitational field equations:
\bsubeq \lab{2.2}
\bea
&&4\nabla_\m\bigl(b\b_i{^{\m\n}}\bigr)
   +4b\b^{nm\n}T_{nmi}-h_i{^\n}b\cL_T =0\, ,             \lab{2.2a}\\
&&\nabla_\m \l_{ij}{^{\m\n}}+2b\b_{[ij]}{^\n}=0\, ,      \lab{2.2b}\\
&&R^{ij}{}_{\m\n}=0 \, .                                 \lab{2.2c}
\eea
\esubeq
Note that the field equations \eq{2.2b} satisfy $6$ differential
identities since the covariant divergence of the left hand side
vanishes on account of $R^{ij}{}_{\m\n}=0$. Hence, the number of
{\it independent\/} equations \eq{2.2b} is $24-6=18$.

The gravitational Lagrangian \eq{2.1} is, by construction, invariant
under the local Poincar\'e transformations:
\bea
&&\d_0 b^k{_\m}=\o^k{_s}b^s{_\m}
        -\xi^\r{}_{,\m}b^k{_\r}-\xi^\r\pd_\r b^k{_\m}\, ,\nn\\
&&\d_0 A^{ij}{_\m}=-\o^{ij}{}_{,\m}
        +\o^i{_s}A^{sj}{_\m}+\o^j{_s}A^{is}{_\m}
        -\xi^\r{}_{,\m}A^{ij}{_\r}-\xi^\r\pd_\r A^{ij}{_\m} \, ,\nn\\
&&\d_0\l_{ij}{}^{\m\n}=\o_i{^s}\l_{sj}{}^{\m\n}+\o_j{^s}\l_{is}{}^{\m\n}
        +\xi^\m{}_{,\r}\l_{ij}{}^{\r\n}+\xi^\n{}_{,\r}\l_{ij}{}^{\m\r}
        -\pd_\r(\xi^\r\l_{ij}{}^{\m\n})\, ,                   \lab{2.3}
\eea
where $\d_0\vphi(x)=\vphi'(x)-\vphi(x)$ is the form variation of
$\vphi(x)$. In addition, it is also invariant, up to a four-divergence,
under the transformations
\bsubeq\lab{2.4}
\be
\d_0\l_{ij}{}^{\m\n}=\nabla_\r\ve_{ij}{}^{\m\n\r} \, ,      \lab{2.4a}
\ee
where the gauge parameter $\ve_{ij}{}^{\m\n\r}=-\ve_{ji}{}^{\m\n\r}$
is completely antisymmetric in its upper indices, and has $6\times 4=24$
components. The invariance is easily verified by using the Bianchi
identity $\nabla_\r R^{ij}{}_{\m\n}+\hbox{cyclic}\,(\m,\n,\r)=0$.
On the other hand, the invariance of the field equation \eq{2.2b}
follows directly from $R^{ij}{}_{\m\n}=0$. The symmetry \eq{2.4a} will
be referred to as $\l$ symmetry.

It is useful to observe that the $\l$ transformations can be written in
the form
\bea
&&\d_0\l_{ij}{}^{\a\b} =\nabla_0\ve_{ij}{}^{\a\b}
         +\nabla_\g\ve_{ij}{}^{\a\b\g}\, ,\qquad
          \ve_{ij}{}^{\a\b}\equiv\ve_{ij}{}^{\a\b 0}\, ,\nn\\
&&\d_0\l_{ij}{}^{0\b}=\nabla_\g\ve_{ij}{}^{\b\g}\, .         \lab{2.4b}
\eea
\esubeq
We shall show in the next section, by canonical methods, that the only
independent parameters of the $\l$ symmetry \eq{2.4b} are
$\ve_{ij}{}^{\a\b}$; in other words, the six parameters
$\ve_{ij}{}^{\a\b\g}$ can be completely discarded. Consequently, the
number of {\it independent\/} gauge parameters is $24-6=18$. They can be
used, for instance, to fix $\l_{ij}{}^{\a\b}$, whereupon the independent
field equations \eq{2.2b} determine $\l_{ij}{}^{0\b}$ (at least locally).

It is evident that the Poincar\'e and $\l$ gauge symmetries are always
present ({\it sure\/} symmetries), independently of the values of
parameters $A,B$ and $C$ in the teleparallel theory \eq{2.1}. Moreover,
it will become clear from the canonical analysis that  there are {\it no
other\/} sure gauge symmetries. Specific models, such as \tgr, may have
{\it extra\/} gauge symmetries,  present only for some special
(critical) values of parameters, but these will not be the subject of
our analysis.
Our task in the subsequent sections is the construction of the sure
gauge generators, describing the Poincar\'e symmetry and the $\l$
symmetry.

\subsub{Hamiltonian and constraints.} Gauge symmetries of a dynamical
system are best described by the related canonical generators. The
program of constructing the gauge generators of the general
teleparallel theory \eq{2.1} demands the complete knowledge of the
Hamiltonian and the constraints \cite{13}. However, the Hamiltonian
structure of the general theory is missing. Instead, we can use the
known Hamiltonian of \tgr, and construct the generators of the $\l$ and
Poincar\'e symmetries in this particular case. We shall see that the
form of the results obtained in this manner has a natural extension to
the general case. After making this extension, the action of the
extended generator on the whole phase space will be explicitly
verified. This approach is essentially based on the ideas used in Ref.
\cite{14} to construct the Poincar\'e gauge generator of the general
$R+T^2+R^2$ theory.

We begin by displaying the Hamiltonian and the constraints of \tgr\
\cite{12}. The canonical Hamiltonian is given in Appendix A.
The general Hamiltonian dynamics is described by the total Hamiltonian:
\bsubeq\lab{2.5}
\bea
\cH_T=\,&&\hat\cH_T+\pd_\a\bar D^\a \, ,\nn\\
\hat\cH_T\equiv\,&&N\bcH_\ort+N^\a\bcH_\a-\fr{1}{2}A^{ij}{_0}\bcH_{ij}
          -\l_{ij}{}^{\a\b}\bcH^{ij}{}_{\a\b}\nn\\
        &&+\, u^i{_0}\p_i{^0}+\fr{1}{2}u^{ij}{_0}\p_{ij}{^0}
 +\fr{1}{4}u_{ij}{}^{\a\b}\p^{ij}{}_{\a\b}+(u\cdot\phi)\,,\lab{2.5a}
\eea
where
\bea
&&\bcH_\ort=\cH_\ort -\fr{1}{8}(\pd\cH_\ort/\pd A^{ij}{}_\a)
                      \pi^{ij}{}_{0\a}\, ,\nn\\
&&\bcH_\a=\pi_i{^\b}T^i{_{\a\b}}-b^k{_\a}\nabla_\b\pi_k{^\b}
     +\fr{1}{2}\pi^{ij}{}_{0\a}\nabla_\b\l_{ij}{}^{0\b}\, ,\nn\\
&&\bcH_{ij}=2\pi_{[i}{^\a}b_{j]\a}+\nabla_\a\pi_{ij}{^\a}
     +2\pi^s{}_{[i0\a}\l_{sj]}{}^{0\a}\, ,\nn\\
&&\bcH^{ij}{}_{\a\b}=R^{ij}{}_{\a\b}
     -\fr{1}{2}\nabla_{[\a}\pi^{ij}{}_{0\b]}\, ,\nn\\
&&\bar D^\a=b^k{_0}\p_k{^\a}+\fr{1}{2}A^{ij}{_0}\p_{ij}{^\a}
   -\fr{1}{2}\l_{ij}{}^{\a\b}\p^{ij}{}_{0\b}\, .         \lab{2.5b}
\eea
\esubeq
The expression for $\cH_\ort$ is defined in Eq. \eq{A.1b}.
Note that $\bcH_\a$ and $\bcH_{ij}$ differ from the corresponding
expressions in Ref. \cite{12} by squares of constraints,
which is irrelevant for our analysis.
The term $(u\cdot\phi)=\fr{1}{2}u^{ik}\bphi_{ik}$ in $\hat\cH_T$
describes extra primary first class constraints specific to
\tgr, which are given in Eq. \eq{A.2}.

The complete dynamical classification of the constraints is given
in the following table:
\begin{center}
\begin{tabular}{|l|l|l|}\hline
\rule{0pt}{12pt} &first class   &second class \\ \hline
\rule[-1pt]{0pt}{15pt}
primary &$\p_i{^0},\p_{ij}{^0},\p^{ij}{}_{\a\b},\bphi_{ik}$
        &$\phi_{ij}{^\a},\p^{ij}{}_{0\b}$  \\ \hline
\rule[-1pt]{0pt}{15pt}
secondary &$\bcH_\ort,\bcH_\a,\bcH_{ij},\bcH^{ij}{}_{\a\b}$
          &\\ \hline
\end{tabular}
\end{center}
where
\be
\phi_{ij}{^\a}=\p_{ij}{^\a}-4\l_{ij}{}^{0\a}\, .          \lab{2.6}
\ee
The constraints $\phi_{ij}{^\a}$ and $\p^{ij}{}_{0\b}$ are
second class since $\{\phi_{ij}{^\a},\p^{kl}{}_{0\b}\}\not\approx 0$.
The first class constraints are identified by observing that they appear
multiplied by arbitrary multipliers or unphysical variables in the total
Hamiltonian \eq{2.5a}.

We display here, for later convenience, the part of the Poisson bracket
algebra of constraints involving $\bcH^{ij}{}_{\a\b}$:
\bea
&&\{\bcH^{ij}{}_{\a\b},\bcH'_{kl}\}=
   (\d^i_k \bcH_l{^j}{}_{\a\b}
   +\d^j_k\bcH^i{_l}{}_{\a\b})\d -(k\lra l)\, ,\nn\\
&&\{\bcH^{ij}{}_{\a\b},\bcH'_\g\}
   =\{\bcH^{ij}{}_{\a\b},\bcH'_\ort\}
   =\{\bcH^{ij}{}_{\a\b},\bcH'^{\,kl}{}_{\g\d}\}= 0 \, .    \lab{2.7}
\eea
The equations $\{\bcH^{ij}{}_{\a\b},\bcH'_\g\}=0$ hold up to squares
of constraints, which are always ignored in an on-shell analysis.

The total Hamiltonian $\cH_T$, in contrast to $\hat\cH_T$, does not
contain the derivatives of momentum variables. The only components of
$\cH_T$ that depend on the specific form of the Lagrangian are
$\bcH_\ort$ and $\bphi_{ij}$. In the next section we shall use the
above canonical structure of \tgr\ to construct the generator of the
$\l$ symmetry \eq{2.4}. The result will be generalized to hold for any
teleparallel theory of the given class \eq{2.1}.

\sect{The \mb{\l} symmetry}

If gauge transformations are given in terms of arbitrary parameters
$\ve(t)$ and their first time derivatives $\dot\ve(t)$, as is the
case with the symmetries of our Lagrangian \eq{2.1}, the gauge
generators have the form
\bsubeq\lab{3.1}
\be
G=\ve(t)\,G^{(0)} + {\dot\ve}(t)\,G^{(1)} \, ,            \lab{3.1a}
\ee
where the phase space functions $G^{(0)}$ and $G^{(1)}$ satisfy the
conditions \cite{13}
\bea
                 G^{(1)}&&=C_{PFC}\, ,\nn\\
G^{(0)}+\{ G^{(1)},H_T\}&&=C_{PFC}\, ,\nn\\
        \{ G^{(0)},H_T\}&&=C_{PFC}\, ,                    \lab{3.1b}
\eea
\esubeq
and $C_{PFC}$ denotes a primary first class (PFC) constraint. These
conditions clearly define the procedure for constructing the generator:
one starts with an arbitrary PFC constraint $G^{(1)}$, evaluates its
Poisson bracket with $H_T$, and defines $G^{(0)}$ in accordance with
$\{G^{(0)},H_T\}=C_{PFC}$.

Now, we are going to construct the gauge generator of the
$\l$ symmetry \eq{2.4}. The only PFC constraint that acts
on the Lagrange multipliers $\l_{ij}{}^{\m\n}$ is $\p_{ij}{}^{\a\b}$.
Starting with $\p^{ij}{}_{\a\b}$ as our $G^{(1)}$, we look for the
generator in the form
\bsubeq\lab{3.2}
\be
G_A(\ve)= \fr{1}{4}\,\dot\ve_{ij}{}^{\a\b}\p^{ij}{}_{\a\b}
       +\fr{1}{4}\,\ve_{ij}{}^{\a\b}S^{ij}{}_{\a\b}\, .    \lab{3.2a}
\ee
The phase space function $S^{ij}{}_{\a\b}$ is to be found from
\eq{3.1b}. In the first step, we obtain the $G^{(0)}$ part of the
generator up to PFC constraints:
$$
S^{ij}{}_{\a\b}= -4\bcH^{ij}{}_{\a\b}+C_{PFC}\, .
$$
Then, using the algebra of constraints involving $\bcH^{ij}{}_{\a\b}$
given in Eq. \eq{2.7}, and the third condition in \eq{3.1b}, we find
\be
S^{ij}{}_{\a\b}= - 4\,\bcH^{ij}{}_{\a\b}
                 + 2 A^{[i}{}_{k0}\,\p^{j]k}{}_{\a\b}\, .  \lab{3.2b}
\ee
\esubeq
This completely defines the generator $G_A(\ve)$ we were looking for.

Applying the generator \eq{3.2} to the fields according to
$\d_0 X = \int d^3x'\{X,G'\}\,$, we find
\be
\d_0^A\l_{ij}{}^{0\a} = \nabla_{\b}\,\ve_{ij}{}^{\a\b}\, ,\qquad
\d_0^A\l_{ij}{}^{\a\b}= \nabla_0\,\ve_{ij}{}^{\a\b} \, ,  \lab{3.3}
\ee
as the only nontrivial field transformations. This result, however,
does not agree with the form of the $\l$ symmetry given in Eq.
\eq{2.4b}, which contains an additional piece,
$\nabla_\g\ve_{ij}{}^{\a\b\g}$, in the expression for
$\d_0\l_{ij}{}^{\a\b}$. Since there are no other PFC constraints
that could produce the transformation of $\l_{ij}{}^{\a\b}$, the
canonical origin of the additional term seems somewhat puzzling.

The solution of the problem is, however, quite simple: if we consider
only independent gauge transformations, this term is not needed,
as it is {\it not independent\/} of what we already have in \eq{3.3}.
To prove this statement, consider the following PFC constraint:
$$
\Pi^{ij}{}_{\a\b\g}=\nabla_\a\p^{ij}{}_{\b\g}
  +\nabla_\g\p^{ij}{}_{\a\b}+\nabla_\b\p^{ij}{}_{\g\a} \, .
$$
This constraint is essentially a linear combination of
$\p^{ij}{}_{\a\b}$; hence, the related gauge generator will not be
truly independent of the above general expression \eq{3.2}.
Further, using the Bianchi identity for $R^{ij}{}_{\a\b}$, one finds
the relation
$$
\nabla_{\a}\bcH^{ij}{}_{\b\g}+\nabla_{\b}\bcH^{ij}{}_{\g\a}+
\nabla_{\g}\bcH^{ij}{}_{\a\b} = 0 \, ,
$$
which holds up to squares of constraints. As a consequence,
$\Pi^{ij}{}_{\a\b\g}$ commutes with the total Hamiltonian up to
PFC constraints, and is, therefore, a correct gauge generator
by itself. Hence, we can introduce a new gauge generator,
\be
G_B(\ve)=-\fr{1}{4}\,\ve_{ij}{}^{\a\b\g}
              \nabla_{\a}\p^{ij}{}_{\b\g}\,,              \lab{3.4}
\ee
where the parameter $\ve_{ij}{}^{\a\b\g}$ is totally antisymmetric
with respect to its upper indices. The only nontrivial field
transformation produced by this generator is
$$
\d_0^B\l_{ij}{}^{\a\b}=\nabla_\g\ve_{ij}{}^{\a\b\g} \, ,
$$
and it coincides with the missing term in Eq. \eq{3.3}. This concludes
the proof that the six parameters $\ve_{ij}{}^{\a\b\g}$ in the $\l$
transformations \eq{2.4} can be completely discarded if we are
interested only in the independent $\l$ transformations.

Although the generator $G_B$ is not truly independent of $G_A$, it is
convenient to define
\be
G(\ve) \equiv G_A(\ve)+G_B(\ve)                             \lab{3.5}
\ee
as an overcomplete gauge generator, since it automatically generates
the covariant Lagrangian form of the $\l$ symmetry, Eq. \eq{2.4}.

The action of the generator \eq{3.5} on momenta is easily seen to be
correct, in the sense that it yields the result in agreement with the
defining relations $\p_A=\pd\cL/\pd\dot\vphi^A$ (see Appendix B). In
particular, the only nontrivial transformation law for the momenta,
$$
\d_0\p_{ij}{}^{\a}=4\nabla_{\b}\,\ve_{ij}{}^{\a\b} \, ,
$$
agrees with \eq{2.4b} through the conservation of the primary
constraint $\phi_{ij}{}^{\a}\approx 0\,$.

The above construction is based on using the first class constraints
$\p^{ij}{}_{\a\b}$, $\bcH^{ij}{}_{\a\b}$, the part of the Poisson
bracket algebra involving these constraints, and the second class
constraints $\phi_{ij}{^\a}$. All these constraints and their
properties are independent of the values of parameters in the theory;
hence, we can conclude that
\bitem
\item[] {\it $G(\ve)$ is the correct generator of $\l$ symmetry in the
general teleparallel theory\/}.
\eitem

\sect{A simple special case}

Before we proceed with the construction of the Poincar\'e gauge
generator, let us make a few comments. The $\cL_T$ part of our
Lagrangian \eq{2.1} is a special case of the general $R+T^2+R^2$
theory, whose Poincar\'e gauge generator has already been constructed
in the literature \cite{14}. In the teleparallel theory of gravity,
this result is to be corrected with the terms stemming from the
$\l_{ij}{}^{\m\n}R^{ij}{}_{\m\n}$ part of the Lagrangian. Since the
general construction procedure is rather complicated, we give in this
section a detailed analysis of the simple special case defined by the
full absence of the torsion part of $\cL$. The obtained results will
provide a clear suggestion for the construction of the Poincar\'e gauge
generator of the general teleparallel theory \eq{2.1}.

The simple Lagrangian we are going to study in this section reads:
\be
\tilde\cL =\l_{ij}{}^{\m\n}R^{ij}{}_{\m\n}\,.               \lab{4.1}
\ee
Clearly, the tetrad variables are absent, but $\tilde\cL$ possesses
all the gauge symmetries of the general teleparallel theory. A
straightforward Hamiltonian analysis gives the total Hamiltonian which
coincides with $\hat\cH_T$, Eq. \eq{2.5}, up to the tetrad related terms:
\bsubeq\lab{4.2}
\be
\tcH_T = -\fr{1}{2}A^{ij}{_0}\tcH_{ij}
   -\l_{ij}{}^{\a\b}\tcH^{ij}{}_{\a\b}+\fr{1}{2}u^{ij}{_0}\p_{ij}{^0}
   +\fr{1}{4}u_{ij}{}^{\a\b}\p^{ij}{}_{\a\b}\,,            \lab{4.2a}
\ee
where
\bea
&&\tcH_{ij}\equiv \nabla_\a\pi_{ij}{^\a}
                      +2\pi^s{}_{[i0\a}\l_{sj]}{}^{0\a}\, ,\nn\\
&&\tcH^{ij}{}_{\a\b}\equiv R^{ij}{}_{\a\b}-\fr{1}{2}
      \nabla_{[\a}\pi^{ij}{}_{0\b]}=\bcH^{ij}{}_{\a\b}\, . \lab{4.2b}
\eea
\esubeq
Among primary constraints, $\p_{ij}{^0}$ and $\p^{ij}{}_{\a\b}$ are
first class, and consequently, responsible for the existence of gauge
symmetries, while $\phi_{ij}{^\a}\equiv \p_{ij}{^\a}-4\l_{ij}{}^{0\a}$
and $\p^{ij}{}_{0\b}$ are second class. The Poisson bracket algebra of
the Hamiltonian constraints has the form
\bea
&&\{\tcH_{ij},\tcH'_{kl}\} =\bigl( \eta_{ik}\tcH_{lj}
   +\eta_{jk}\tcH_{il} \bigr)\d -(k\lra l)\,, \nn\\
&&\{\tcH^{ij}{}_{\a\b},\tcH'_{kl}\} = (\d^i_k \tcH_l{^j}{}_{\a\b}
   +\d^j_k\tcH^i{_l}{}_{\a\b})\d -(k\lra l)\, ,      \nn\\
&&\{\tcH^{ij}{}_{\a\b},\tcH'^{\,kl}{}_{\g\d}\}=0 \,.       \lab{4.3}
\eea

\subsub{The Lorentz generator}. We begin with the PFC constraint
$\p_{ij}{^0}$, and define
\bsubeq\lab{4.4}
\be
\tilde G(\o) = -\fr{1}{2}\dot\o^{ij}\p_{ij}{^0}
               -\fr{1}{2}\o^{ij}\tilde S_{ij}\, .         \lab{4.4a}
\ee
From the second condition in \eq{3.1b}, we obtain
$\tilde S_{ij}=-\tcH_{ij} + C_{PFC}$. Then, using the constraint
algebra \eq{4.3} and the third condition in \eq{3.1b}, we find the
complete function $\tilde S_{ij}$ to read:
\be
\tilde S_{ij}=-\tcH_{ij}+2A^s{}_{[i0}\p_{sj]}{^0}
              +\l_{s[i}{}^{\a\b}\p^s{}_{j]\a\b} \,.       \lab{4.4b}
\ee
\esubeq
It is easy to verify that the action of the Lorentz gauge generator
$\tilde G(\o)$ on the fields $A^{ij}{_\m}$ and $\l_{ij}{}^{\m\n}$ has
the expected form, coinciding with the $\o$ piece of the Poincar\'e
transformations \eq{2.3}.

\subsub{The \mb{\l} generator}. The $\l$ gauge generator is obtained
by starting Castellani's procedure \eq{3.1} with the PFC constraint
$\p^{ij}{}_{\a\b}\,$. All the steps of the construction, the analysis
and the final result are the same as in Sect. 3. Thus,
$\tilde G(\ve)=G(\ve)\,$, and we rewrite it as
\be
\tilde G(\ve) = \fr{1}{4}\,\dot\ve_{ij}{}^{0\a\b}\p^{ij}{}_{\a\b}
 +\fr{1}{4}\,\ve_{ij}{}^{0\a\b}S^{ij}{}_{\a\b}
 -\fr{1}{4}\,\ve_{ij}{}^{\a\b\g}\nabla_{\a}\p^{ij}{}_{\b\g}
                                                     \,, \lab{4.5}
\ee
with $S^{ij}{}_{\a\b}$ given by Eq. \eq{3.2b}. The action on the
fields is the same as in Eq. \eq{2.4}.

Therefore, the action of the complete generator
$\tilde G(\o)+\tilde G(\ve)$ gives:
\bea
\d_0 A^{ij}{}_{\m}
    =&&\ \o^i{_k}\,A^{kj}{}_{\m}+\o^j{_k}\,A^{ik}{}_{\m}
      -\o^{ij}{}_{,\,\m}  \nn\\
\d_0\l_{ij}{}^{\m\n}=
     &&\ \o_i{^k}\,\l_{kj}{}^{\m\n}+\o_j{^k}\,\l_{ik}{}^{\m\n}
       +\nabla_{\l}\,\ve_{ij}{}^{\m\n\l}\,.               \lab{4.6}
\eea
The transformation laws \eq{4.6} exhaust the gauge symmetries of the
simple theory \eq{4.1}. Notice, however, that the Lagrangian
$\tilde\cL$ also possesses the local translational symmetry, which
has not been obtained by Castellani's procedure. If Castellani's
algorithm is an exhaustive one, then the translational symmetry
must be somehow hidden in the above result \eq{4.6}.
In what follows, we shall demonstrate that this is really true, namely
that the translational symmetry emerges from a simple redefinition of
the gauge parameters in \eq{4.6}.

\subsub{The Poincar\'e generator}. Let us consider the following
replacement of the parameters $\o^{ij}$ and $\ve_{ij}{}^{\m\n\l}$ in
Eq. \eq{4.6}:
\be
\o^{ij} \ra \o^{ij}+\xi^{\m}A^{ij}{}_{\m}\, ,\qquad
   \ve_{ij}{}^{\m\n\l} \ra
  -\bigl(\xi^{\m}\l_{ij}{}^{\n\l}+\xi^{\n}\l_{ij}{}^{\l\m}
  +\xi^{\l}\l_{ij}{}^{\m\n}\bigr) \,.                       \lab{4.7}
\ee
The resulting on-shell field transformations,
\bea
\d_0 A^{ij}{}_{\m}\approx \
&& \o^i{_k}\,A^{kj}{}_{\m}+\o^j{_k}\,A^{ik}{}_{\m}
  -\o^{ij}{}_{,\,\m}-\xi^{\n}{}_{,\,\m}\,A^{ij}{}_{\n}-
   \xi^{\n}A^{ij}{}_{\m ,\,\n}\,, \nn\\
\d_0 \l_{ij}{}^{\m\n}\approx \
&& \o_i{^k}\,\l_{kj}{}^{\m\n}+\o_j{^k}\,\l_{ik}{}^{\m\n}
  +\xi^{\m}{}_{,\,\r}\,\l_{ij}{}^{\r\n}
  +\xi^{\n}{}_{,\,\r}\,\l_{ij}{}^{\m\r}
  -\pd_{\r}\bigl(\xi^{\r}\,\l_{ij}{}^{\m\n}\bigr)\,,\nn
\eea
are the exact Poincar\'e gauge transformations we expected to
find in this theory. As we can see, the local translations are not
obtained as independent gauge transformations, but rather emerge as
a part of the $\l$ and Lorentz symmetries in \eq{4.6}. The corresponding
Poincar\'e generator is obtained by using the replacement \eq{4.7}
in the gauge generator $\tilde G(\o)+\tilde G(\ve)$. Thus, we find:
\bsubeq\lab{4.8}
\be
\tilde G=\tilde G(\o)+\tilde G(\xi)\, ,                   \lab{4.8a}
\ee
where the first term, describing local Lorentz rotations, has the
form \eq{4.4}, while the second term, describing local translations,
is given by
\bea
\tilde G(\xi) =
&&  -\dot\xi^0\bigl( \fr{1}{2}A^{ij}{_0}\p_{ij}{^0}
    +\fr{1}{4}\l_{ij}{}^{\a\b}\p^{ij}{}_{\a\b} \bigr)
    -\xi^0\,\tcH_T  \nn\\
&&-\dot\xi^\a\bigl( \fr{1}{2}A^{ij}{_\a}\p_{ij}{^0}
    -\fr{1}{2}\l_{ij}{}^{0\b}\p^{ij}{}_{\a\b}\bigr) \nn\\
&&-\xi^\a\bigl[ \tilde\cP_\a
   -\fr{1}{4}\l_{ij}{}^{\b\g}\pd_\a\p^{ij}{}_{\b\g}
   -\fr{1}{2}\pd_\g\bigl(\l_{ij}{}^{\b\g}\p^{ij}{}_{\a\b}
    \bigr)\bigr] \, .                                      \lab{4.8b}
\eea
In the above expressions we used the following notation:
\bea
&&\tilde\cP_\a \equiv \tcH_\a-\fr{1}{2}A^{ij}{_\a}\tcH_{ij}
  +2\l_{ij}{}^{0\b}\tcH^{ij}{}_{\a\b}
  +\fr{1}{2}\p_{ij}{^0}\pd_\a A^{ij}{_0}\,, \nn\\
&&\tcH_\a \equiv \fr{1}{2}\pi^{ij}{}_{0\a}
                      \nabla_\b\l_{ij}{}^{0\b}\,.         \lab{4.8c}
\eea
\esubeq
Notice that the term $\tcH_\a$ in $\tilde G(\xi)$ has the structure
of squares of constraints, and therefore, does not contribute to the
nontrivial field transformations. Nevertheless, we shall retain it in
the generator because it makes the field transformations practically
off shell (up to $R^{ij}{}_{\a\b}\approx 0$). This will help us to
straightforwardly find the form of the extension of $\tilde G$ in the
general teleparallel theory.

\sect{Poincar\'e gauge symmetry}

Staring from the Poincar\'e gauge generator \eq{4.8} of the simple
theory \eq{4.1}, and comparing it with the earlier results obtained in
Ref. \cite{14}, it is almost evident how its modification to include
the tetrad sector should be defined. In this section, we are going to
prove that the complete Poincar\'e gauge generator  of the general
teleparallel theory \eq{2.1} has the form
\bsubeq\lab{5.1}
\be
G=G(\o)+G(\xi)\, ,                                         \lab{5.1a}
\ee
where the first term describes local Lorentz rotations,
\be
G(\o)=-\fr{1}{2}\dot\o^{ij}\p_{ij}{^0}-\fr{1}{2}\o^{ij}S_{ij}\, ,
                                                           \lab{5.1b}
\ee
while the second term describes local translations,
\bea
G(\xi)=&&\,-\dot\xi^0\bigl(
   b^k{_0}\p_k{^0}+\fr{1}{2}A^{ij}{_0}\p_{ij}{^0}
  +\fr{1}{4}\l_{ij}{}^{\a\b}\p^{ij}{}_{\a\b} \bigr)-\xi^0\cP_0 \nn\\
&&-\dot\xi^\a\bigl(
   b^k{_\a}\p_k{^0}+\fr{1}{2}A^{ij}{_\a}\p_{ij}{^0}
  -\fr{1}{2}\l_{ij}{}^{0\b}\p^{ij}{}_{\a\b}\bigr)\nn\\
&&-\xi^\a\bigl[ \bar\cP_\a
  -\fr{1}{4}\l_{ij}{}^{\b\g}\pd_\a\p^{ij}{}_{\b\g}
  -\fr{1}{2}\pd_\g\bigl(\l_{ij}{}^{\b\g}\p^{ij}{}_{\a\b}\bigr)\bigr]
                                                     \,.  \lab{5.1c}
\eea
In the above expressions, we used the following notation:
\bea
&&S_{ij}=-\bcH_{ij}+ 2b_{[i0}\p_{j]}{^0}
 +2A^s{}_{[i0}\p_{sj]}{^0}+2\l_{s[i}{}^{\a\b}\p^s{}_{j]\a\b}\, ,\nn\\
&&\cP_0\equiv \hat\cH_T=\cH_T-\pd_\a\bar D^\a\, ,\nn\\
&&\bar\cP_\a= \bcH_\a-\fr{1}{2}A^{ij}{_\a}\bcH_{ij}
 +2\l_{ij}{}^{0\b}\bcH^{ij}{}_{\a\b} +\p_k{^0}\pd_\a b^k{_0}
 +\fr{1}{2}\p_{ij}{^0}\pd_\a A^{ij}{_0}\, .               \lab{5.1d}
\eea
\esubeq
The form of the total Hamiltonian $\cH_T$ is defined by the choice of
the Lagrangian: in the case of \tgr, it is determined by Eq. \eq{2.5a},
while the general $\cH_T$ is constructed by the principles of Appendix
C. The Poincar\'e generator $G$ is obtained from the simplified
expression $\tilde G$ in Eq. \eq{4.8a} by a natural process of
extension, which consists of
\bitem
\item[]\bull the replacements
   $\tcH_\a\to\bcH_\a$, $\tcH_{ij}\to\bcH_{ij}$,
   $\tcH^{ij}{}_{\a\b}\to\bcH^{ij}{}_{\a\b}$,  \\
\bull the addition of $-\o^{ij}b_{[i}{^0}\p_{j]0}$,
   $-\dot\xi^0 b^k{_0}\p_k{^0}$, $-\dot\xi^\a b^k{_\a}\p_k{^0}$,
   $-\xi^\a\p_k{^0}\pd_\a b^k{_0}$, and \\
\bull the replacement $\tilde\cH_T\to\hat\cH_T$.
\eitem
This amounts to completing the Poincar\'e gauge generator so as to act
correctly also in the tetrad sector \cite{14}.

The proof that the Poincar\'e gauge generator has the form \eq{5.1} is
realized by showing that $G$ produces the correct Poincar\'e gauge
transformations on the complete phase space, i.e. on all the fields and
momenta.

\subsub{Action on the fields.} We now demonstrate that the action of
the generator \eq{5.1} on the fields produces the complete Poincar\'e
gauge transformations \eq{2.3}.

It is straightforward to verify $\o^{ij}$ and $\xi^\a$ transformations
in \eq{2.3}. The derivation of $\xi^0$ transformations is more subtle.
Let us illustrate the procedure on $\l_{ij}{}^{0\b}$:
\bea
\d_0(\xi^0)\l_{ij}{}^{0\b}\,&&=
 -\int d^3x'\,\xi'{^0}\{\l_{ij}{}^{0\b},\cP'_0\} \nn\\
&&=-\int d^3x'\,\xi'{^0}\{\l_{ij}{}^{0\b},\cH'_T-\pd'_\g\bar D'{^\g}\}
 \approx-\xi^0\dot\l_{ij}{}^{0\b}+(\pd_\g\xi^0)\l_{ij}{}^{\g\b}\, .\nn
\eea
In deriving the last (weak) equality we used the relation
$\int d^3x'\,\xi'{^0}\{\l_{ij}{}^{0\b},\cH'_T\}
   \approx\xi^0\dot\l_{ij}{}^{0\b}$, which is based on the fact that
$\cH_T$ does not depend on the derivatives of momentum variables.
For \tgr\ this is verified by an explicit inspection of $\cH_T$, while
for the general case we use the arguments of Appendix C.

In a similar way, one can find the (on-shell) transformation rules for
the other fields, and they all agree with Eq. \eq{2.3}. The
calculations are based on using the Hamiltonian equations of motion and
the sure constraints $R^{ij}{}_{\a\b}\approx 0$.

To summarize, the only properties used in the derivation are:
\bitem
\item[]\bull $\cH_T$ does not depend on the derivatives of momenta, \\
\bull it governs the time evolution of dynamical variables
      by $\dot Q=\{ Q,H_T\}$, \\
\bull $R^{ij}{}_{\a\b}\approx 0$ are the sure constraints of the theory.
\eitem
Consequently, the obtained transformation rules of the fields are
correct for an arbitrary choice of parameters in the general
teleparallel theory \eq{2.1}.

\subsub{Action on the momenta.} In the next step, we are going to
compare the action of the generator \eq{5.1} on the momenta
$(\p_k{^\m},\p_{ij}{^\m},\p^{ij}{}_{\m\n})$ with the correct
transformation rules for these variables. The correct rules
are determined by the defining relation $\p_A=\pd\cL/\pd\dot\vphi^A$,
and the known transformation laws for the fields. The general formula
derived in Appendix B leads to
\bea
&&\d\p_k{^\m}=\o_k{^s}\p_s{^\m} + \xi^\m{}_{,\r}\p_k{^\r}
  +\xi^0{}_{,\g}{\pd\cL\over\pd b^k{}_{\m,\g}}
  -\xi^\g{}_{,\g}\p_k{^\m}      \, , \nn\\
&&\d\p_{ij}{^\m}=\o_i{^s}\p_{sj}{^\m}+\o_j{^s}\p_{is}{^\m}
  +\xi^\m{}_{,\r}\p_{ij}{^\r}
  +\xi^0{}_{,\g}{\pd\cL\over\pd A^{ij}{}_{\m,\g}}
  -\xi^\g{}_{,\g}\p_{ij}{^\m}   \, , \nn\\
&&\d\p^{ij}{}_{\m\n}=
   \o^i{_s}\p^{sj}{}_{\m\n}+\o^j{_s}\p^{is}{}_{\m\n}
  -\xi^\r{}_{,\m}\p^{ij}{}_{\r\n}- \xi^\r{}_{,\n}\p^{ij}{}_{\m\r}
  +\xi^0{}_{,0}\p^{ij}{}_{\m\n}\, .                      \lab{5.2}
\eea
To check if the generator \eq{5.1} produces the
above gauge transformations of momenta, we shall use
the results of Appendix C, and the relations $\phi_{ij}{^\a}\approx 0$,
$\pd\cL/\pd A^{ij}{}_{\m,\a}=-4\l_{ij}{}^{\m\a}$, which characterize the
teleparallel theory \eq{2.1} for any choice of the parameters $A,B,C$.

We begin by noting that the generator \eq{5.1} has the standard form
in the tetrad sector; hence, it follows that the transformation law
for $\p_k{^\m}$ has the correct form given in Eq. \eq{5.2}, as has
been shown in Ref. \cite{14}.

Since all $\o^{ij}$ transformations can be verified
straightforwardly, we focus our attention on $\xi^\m$
transformations. Consider, first, the $\xi^\m$ transformations of
$\p_{ij}{^\a}$:
\bea
\d_0(\xi^0)\p_{ij}{^\a}&&=\int d^3x'\{\p_{ij}{^\a},-(\xi^0\cP_0)'\}
   =-\xi^0\dot\p_{ij}{^\a}-4\xi^0{}_{,\g}\l_{ij}{}^{\a\g}\, ,\nn\\
\d_0(\xi^\g)\p_{ij}{^\a}&&= \int d^3x'\{\p_{ij}{^\a},
   -(\fr{1}{2}\dot\xi^\g A^{mn}{_\g}\p_{mn}{^0}+\xi^\b\cP_\b)'\}\nn\\
  &&\approx\dot\xi^\a\p_{ij}{^0}-\pd_\g(\xi^\g\p_{ij}{}^\a)
   +\xi^\a{}_{,\g}\p_{ij}{}^\g\, .\nn
\eea
Here, the last (weak) equality is obtained by discarding terms
proportional to $\phi_{ij}{^\a}$. Comparing with Eq. \eq{5.2},
we find the complete agreement. Note that these transformations,
combined with those in Eq. \eq{2.3}, lead to
$\d_0\phi_{ij}{^\a}\approx 0$, as they should.

In a similar manner, one can show that the $\xi^\m$ transformations
of $\p_{ij}{^0}$ and $\p^{ij}{}_{\m\n}$ agree with those displayed
in Eq. \eq{5.2}. In the process of demonstrating this property we
have used only those relations that characterize an arbitrary
teleparallel theory. Hence,
\bitem
\item[] {\it the expression \eq{5.1} is the correct generator of
Poincar\'e gauge transformations for any choice of parameters in the
teleparallel theory \eq{2.1}\/}.
\eitem

\sect{Concluding remarks}

Gauge structure of the general teleparallel theory \eq{2.1} is
characterized by some specific features, as compared to the standard
PGT \cite{14}. We found two types of sure gauge symmetries, which are
always present in the theory, independently of the values of parameters
in Eq. \eq{2.1}.

The first type is the so-called $\l$ symmetry \eq{2.4}, with 24
Lagrangian parameters. The related canonical generator \eq{3.2} is
based on the PFC constraints $\p^{ij}{}_{\a\b}$. The number of
independent parameters of the $\l$ symmetry is shown to be not 24, but
only 18, which clarifies the true dynamical meaning of the covariant
Lagrangian symmetry. They can be used to fix 18 Lagrange multipliers,
while the remaining 18 can be found using the same number of
independent field equations \eq{2.2b}.

The second type is the usual Poincar\'e gauge symmetry \eq{2.3}.
Although the meaning of this symmetry is well known, the construction
of the related gauge generator shows some unusual features. In
particular, we have found in the simple model of Sect. 4 that local
translations are not obtained as independent gauge transformations, but
rather emerge as a part of the Lorentz and $\l$ symmetries.

The Hamiltonian analysis of the present paper gives a very clear
picture of the general gauge structure of the teleparallel theory,
which has been the subject of many discussions in the past \cite{8,9}.
The two gauge symmetries completely describe the gauge structure of the
theory \eq{2.1}, in the sense that there are no other sure gauge
symmetries. The canonical gauge generators obtained here will be very
useful in studying the important problem of the conservation laws of
energy, momentum and angular momentum \cite{11,12}.

\ack{Acknowledgments}

This work was partially supported by the Serbian Science Foundation,
Yugoslavia.

\appendix

\app{Canonical description of \tgr}

In this Appendix, we present some formulas related to the canonical
description of \tgr.
The canonical Hamiltonian density of \tgr\ can be written in the
form \cite{12}
\bsubeq\lab{A.1}
\bea
\cH_c= N\cH_\ort+N^\a\cH_\a-\fr{1}{2}A^{ij}{_0}\cH_{ij}
      -\l_{ij}{}^{\a\b}R^{ij}{}_{\a\b} +\pd_\a D^\a \, , \lab{A.1a}
\eea
where
\bea
&&\cH_{ij}=2\pi_{[i}{^\a}b_{j]\a}+\nabla_\a\pi_{ij}{^\a} \, ,\nn\\
&&\cH_\a=\pi_i{^\b}T^i{_{\a\b}}-b^k{_\a}\nabla_\b\pi_k{^\b}\, ,\nn\\
&&\cH_\ort=\fr{1}{2}P_T^2-J\bcL_T(\bT)-n^k\nabla_\b\p_k{^\b}\,,\nn\\
&&D^\a=b^k{_0}\p_k{^\a}+\fr{1}{2}A^{ij}{_0}\p_{ij}{^\a}\,,\lab{A.1b}
\eea
and
\bea
P_T^2\,&&={1\over 2aJ}\left( \hp_{(\bi\bk)}\hp^{(\bi\bk)}
         -{1\over 2}\hp^\bm{_\bm}\hp^\bn{_\bn} \right)\, ,\nn\\
\bcL_T(\bT)\,&&=a\left(\fr{1}{4}T_{m\bn\bk}T^{m\bn\bk}
         +\fr{1}{2}T_{\bm\bn\bk}T^{\bn\bm\bk}
         -T^\bm{}_{\bm\bk}T_\bn{}^{\bn\bk} \right) \, .   \lab{A.1c}
\eea
\esubeq
Here, $\nabla_k=h_k{^\m}\nabla_\m$ is the covariant derivative,
$n_k=h_k{^0}/\sqrt{g^{00}}$ is the unit normal to the hypersurface
$x^0=$ const, the bar over the Latin index is defined by the
decomposition
$$
V_k=V_\ort n_k+V_\bk\, ,\qquad V_\ort=n^kV_k\, ,
$$
of an arbitrary vector $V_k$, $\hp_{i\bk}=\p_i{^\a}b_{k\a}$, $N$ and
$N^\a$ are lapse and shift functions, respectively, $N=n_kb^k{_0}$,
$N^\a=h_\bk{^\a}b^k{_0}$, and $J$ is defined by $b=NJ$.
Note that $\cH_c$ is linear in unphysical variables
$(b^k{_0},A^{ij}{_0},\l_{ij}{}^{\a\b})$.

A specific feature of \tgr\ is the existence of the extra PFC
constraints $\bphi_{ij}$ \cite{12}, which appear in the total
Hamiltonian \eq{2.5a}:
\bea
&&\bphi_{ik}=\phi_{ik}-\fr{1}{4}a\bigl(
  \pi_i{^s}{}_{0\a}B_{sk}^{0\a}+\pi_k{^s}{}_{0\a}B_{is}^{0\a}\bigr)
                                                          \, ,\nn\\
&&\phi_{ik}=\hp_{i\bk}-\hp_{k\bi}+a\nabla_\a B^{0\a}_{ik}\, ,\qquad
 B^{0\a}_{ik}\equiv\ve^{0\a\b\g}\ve_{ikmn}b^m_\b b^n_\g\, . \lab{A.2}
\eea

\app{General transformation laws for momenta}

To find the correct transformation rules for momentum variables
$\p_A=\pd\cL/\pd\dot\vphi^A$ with respect to the spacetime
transformations
$$
x'^\m=x^\m+\xi^\m\,,\qquad \vphi{'}^A(x')=\vphi^A(x)+\d\vphi^A(x)\,,
$$
we assume that the Lagrangian $\cL(\vphi^A,\pd\vphi^A)$ is a scalar
density,
$$
\d_0\cL +\pd_\m(\xi^\m\cL)=0\, .
$$
Then, it follows that the momenta $\p_A$ transform in the following way:
\be
\d\p_A=-\p_B{\pd\d\vphi^B\over\pd\vphi^A}
  +\xi^0{}_{,\a}{\pd\cL\over\pd\vphi^A{}_{,\a}}
  -\xi^\a{}_{,\a}\p_A\, ,                                  \lab{B.1}
\ee
where $\d=\d_0+\xi^\r\pd_\r$.

Applying this general formula to the Poincar\'e gauge transformations
\eq{2.3}, we obtain the transformation laws \eq{5.2} for
$(\p_k{^\m},\p_{ij}{^\m},\p^{ij}{}_{\m\n})$.

\app{Dependence of \mb{\cH_T} on fields and momenta}

Let us consider Lagrangians $\cL(\vphi^A,\pd\vphi^A)$ that are at
most quadratic in the first field derivatives. In this case, the
canonical momenta $\p_A=\pd\cL/\pd\dot\vphi^A$ are functions linear
in $\dot\vphi^A$. If the canonical Hamiltonian is constructed in the
standard way, $\cH_c =\p_A \dot\vphi^A -\cL$, the corresponding
total Hamiltonian $\cH_T =\cH_c +u^k\phi_k$ can be written in the
same way:
\bsubeq\lab{C.1}
\be
\cH_T = \p_A \dot\vphi^A - \cL \ ,                         \lab{C.1a}
\ee
where, now, the velocities $\dot\vphi^A$ are functions not only of
the fields and momenta, but also of the multipliers,
\be
\dot\vphi^A = \dot\vphi^A (\vphi , \p , u) \ .             \lab{C.1b}
\ee
\esubeq
The velocities $\dot\vphi^A$ can depend on momentum derivatives only
through the determined multipliers.

Now, using Eq. \eq{C.1}, we find how $\cH_T$ depends on momentum
derivatives:
\be
\frac{\pd\cH_T}{\pd\p_{A,\,\a}} =
  \left( \p_B-\frac{\pd\cL}{\pd\dot\vphi^B} \right)
  \frac{\pd\dot\vphi^B}{\pd\p_{A,\,\a}} \approx 0 \ .       \lab{C.2}
\ee
In a similar way, one can see that $\cH_T$ does not depend on higher
derivatives of momenta, either.

As for the dependence of $\cH_T$ on fields, we can write
\be
\frac{\pd\cH_T}{\pd\vphi^A} = -\frac{\pd\cL}{\pd\vphi^A}
 +\left( \p_B-\frac{\pd\cL}{\pd\dot\vphi^B} \right)
  \frac{\pd\dot\vphi^B}{\pd\vphi^A}
  \approx -\frac{\pd\cL}{\pd\vphi^A} \ ,                    \lab{C.3}
\ee
and similarly for the field derivatives,
\be
\frac{\pd\cH_T}{\pd\vphi^A{}_{,\,\a}} \approx
  -\frac{\pd\cL}{\pd\vphi^A{}_{,\,\a}} \ , \qquad
\frac{\pd\cH_T}{\pd\vphi^A{}_{,\,\a\b}} \approx
  -\frac{\pd\cL}{\pd\vphi^A{}_{,\,\a\b}}\ .                 \lab{C.4}
\ee
Notice that the equalities \eq{C.3} and \eq{C.4} are on-shell
equalities. Our $\cH_T$ in \eq{2.5} is, by construction, of the same
type as $\cH_T$ in \eq{C.1}, and consequently, satisfies the above
relations.

\end{document}